\documentclass[aps,prd,onecolumn,showpacs,nofootinbib,amsmath,amssymb,floatfix,superscriptaddress,showkeys]{revtex4}
\usepackage{graphicx}
\usepackage{dcolumn}
\usepackage{bm}
\usepackage{captcont}
\usepackage{xcolor}
\usepackage{threeparttable}

\usepackage{epstopdf}
\usepackage{mathtools}
\usepackage{natbib}
\usepackage{ulem}
\usepackage{color,txfonts}
\usepackage{hyperref}
\usepackage{subfigure}

\def\ApJ{Astrophys. J.~}
\def\ApJL{Astrophys. J. Lett.~}
\def\AAA{Astron. Astrophys.~}
\def\MNRAS{Mon. Not. R. Astron. Soc.~}
\def\ARAA{Ann. Rev. Astron. Astrophys.~}
\def\PhRvD{Phys. Rev. D~}
\def\etal{{\it et al.~}}

\begin{document}

\title{Estimating the maximum gravitational mass of nonrotating neutron stars from the GW170817/GRB 170817A/AT2017gfo observation}

\author{Dong-Sheng Shao}
\affiliation{Key Laboratory of Dark Matter and Space Astronomy, Purple Mountain Observatory, Chinese Academy of Sciences, Nanjing 210033, China}
\affiliation{School of Astronomy and Space Science, University of Science and Technology of China, Hefei, Anhui 230026, China}
\author{Shao-Peng Tang}
\affiliation{Key Laboratory of Dark Matter and Space Astronomy, Purple Mountain Observatory, Chinese Academy of Sciences, Nanjing 210033, China}
\affiliation{School of Astronomy and Space Science, University of Science and Technology of China, Hefei, Anhui 230026, China}
\author{Xin Sheng}
\affiliation{College of Creative Studies, University of California, Santa Barbara, California 93106, USA}
\author{Jin-Liang Jiang}
\affiliation{Key Laboratory of Dark Matter and Space Astronomy, Purple Mountain Observatory, Chinese Academy of Sciences, Nanjing 210033, China}
\affiliation{School of Astronomy and Space Science, University of Science and Technology of China, Hefei, Anhui 230026, China}
\author{Yuan-Zhu Wang}
\affiliation{Key Laboratory of Dark Matter and Space Astronomy, Purple Mountain Observatory, Chinese Academy of Sciences, Nanjing 210033, China}
\author{Zhi-Ping Jin}
\affiliation{Key Laboratory of Dark Matter and Space Astronomy, Purple Mountain Observatory, Chinese Academy of Sciences, Nanjing 210033, China}
\affiliation{School of Astronomy and Space Science, University of Science and Technology of China, Hefei, Anhui 230026, China}
\author{Yi-Zhong Fan}
\email[Corresponding author.~]{yzfan@pmo.ac.cn}
\affiliation{Key Laboratory of Dark Matter and Space Astronomy, Purple Mountain Observatory, Chinese Academy of Sciences, Nanjing 210033, China}
\affiliation{School of Astronomy and Space Science, University of Science and Technology of China, Hefei, Anhui 230026, China}
\author{Da-Ming Wei}
\affiliation{Key Laboratory of Dark Matter and Space Astronomy, Purple Mountain Observatory, Chinese Academy of Sciences, Nanjing 210033, China}
\affiliation{School of Astronomy and Space Science, University of Science and Technology of China, Hefei, Anhui 230026, China}

\begin{abstract}
Assuming that the differential rotation of the massive neutron star (NS) formed in the double NS (DNS) mergers has been effectively terminated by the magnetic braking and a uniform rotation has been subsequently established (i.e., a supramassive NS is formed), we analytically derive in this work an approximated expression for the critical total gravitational mass ($M_{\rm tot,c}$) to form supramassive NS (SMNS) in the DNS mergers, benefited from some equation of state (EoS) insensitive relationships. The maximum gravitational mass of the nonrotating NSs ($M_{\rm TOV}$) as well as the dimensionless angular momentum of the remnant ($j$) play the dominant roles in modifying $M_{\rm tot,c}$, while the radius and mass differences of the premerger NSs do not. The GW170817/GRB 170817A/AT2017gfo observations have provided so far the best opportunity to quantitatively evaluate $M_{\rm TOV}$. Supposing the central engine for GRB 170817A is a black hole quickly formed in the collapse of an SMNS, we find $M_{\rm TOV}=2.13^{+0.09}_{-0.08}M_\odot$ (68.3\% credibility interval, including also the uncertainties of the EoS insensitive relationships), which is consistent with the constraints set by current NS mass measurements.
\end{abstract}


\maketitle

\section{Introduction} \label{sec:intro}

The maximum gravitational mass of nonrotating neutron star ($M_{\rm TOV}$) is predicted theoretically by the well-known Tolman-Oppenheimer-Volkoff equations together with the equation of state (EoS) of the ultradense matter \citep{Oppenheimer1939,Zwicky1939}.
An accurate knowledge of $M_{\rm TOV}$ plays an important role not only in probing the properties and interactions of cold, ultradense matter but also in understanding many astrophysical phenomena \citep{Lattimer2012}.

So far, the actual value of $M_{\rm TOV}$ is still uncertain. Theoretically, on the basis of general relativity, the principle of causality and Le Chatelier's principle, \citet{Rhoades1974} suggested $M_{\rm TOV}\leq 3.2M_\odot$. Within the same framework, a tighter upper bound $M_{\rm TOV}\leq 2.9M_\odot$ has been set with a more advanced equation of states \citep{Kalogera1996}. \citet{Akmal1998} considered the possibility that matter is maximally incompressible above an assumed density, and showed that realistic models of nuclear forces limit $M_{\rm TOV}\leq 2.5M_\odot$. Anyhow, looser bounds such as $M_{\rm TOV}\leq 4.8M_\odot$ have also been suggested (e.g.,\citep{Hartle1978}). Observationally, there are more than 2000 neutron stars detected in the Galaxy and the masses of a small fraction of these objects, usually in either pulsar-white dwarf or double neutron star binary systems, have been accurately measured. Among these compact object binary systems, the millisecond pulsar (MSP)$-$white dwarf binary systems were found to be ideal candidates in searching for massive neutron stars because MSPs are usually recycled and massive, and the Shapiro delay in this system tends to be detected easily and accurately for the very short pulsar timing. Indeed, records of the most massive neutron star was broken over and over by MSP$-$white dwarf systems, once $1.93\pm0.07M_{\odot}$ (PSR J1614-2230\citep{Demorest2010}), then $2.01\pm0.04M_{\odot}$ (PSR J0348+0432\citep{Antoniadis2013}), and now $2.14^{+0.10}_{-0.09}M_{\odot}$ (PSR J0740+6620\citep{Cromartie2019}). Consequently, we have $M_{\rm TOV}>1.97M_\odot$ or more optimistically $>2.04M_\odot$ (This value is slightly lower than the 68.3\% lower limit set by PSR J0740+6620 because this MSP rotates so quickly that it can enhance $M_{\rm max}$ beyond $M_{\rm TOV}$ by $\sim 0.01M_\odot$.)

Other than the above physical and observational approaches, one can also estimate $M_{\rm TOV}$ with the information from short gamma-ray bursts (sGRBs), a kind of brief $\gamma$-ray flashes widely believed to be powered by the mergers of double neutron stars (DNSs)\citep{Eichler1989}.
The remnant formed in a DNS merger, in principle, could be a promptly formed black hole, a hypermassive NS (HMNS) supported by the differential rotation, a supramassive NS (SMNS) supported by its rapid uniform rotation, or even a stable NS, depending mainly on the EoS models and the masses of the premerger DNSs (e.g.,\citep{Duncan1992, Davies1994, Katz1997, Dai1998, Shibata2000, Baumgarte1999, Morrison2004,Hotokezaka2013a,Kastaun2013,Shibata2015,Foucart2016,Ciolfi2017,Radice2018}). Some afterglows of sGRBs are characterized by extended peculiar x-ray emissions (followed by abrupt cutoffs), which indicate the prolonged activities of the central engines. One hypothetical interpretation is that the central engines were magnetized SMNSs, which did not collapse until they lost most of their rotational kinetic energies \citep{GaoFan2006,FanXu2006,Metzger2008}. Within the SMNS model for the peculiar x-ray afterglow data of sGRBs, a $M_{\rm TOV}\sim 2.2-2.3M_\odot$ was inferred in the literature (e.g.,\citep{Fan2013a,Lasky2014,Li2014,Lu2014,Fryer2015,Lawrence2015,Gao2016}).
The measurements of the first DNS merger gravitational wave event GW170817 \citep{Abbott2017a}, the associated sGRB 170817A \citep{Goldstein2017}, and the follow-up multiwavelength emission \citep{Abbott2017b} have provided a nice opportunity to estimate $M_{\rm TOV}$. An upper bound of $M_{\rm TOV}$ was set to $\sim 2.17M_{\odot}$ if a black hole was formed quickly, though not promptly from the collapse of the neutron star rotating at the mass shedding limit, after the DNS merger \citep{Margalit2017,Shibata2017,Rezzolla2018,Ma2018,Ruiz2018}.
Such a bound would be substantially loosened if the precollapse remnant formed in the merger did not collapse until a good fraction of its initial angular momentum had been lost (see Fig.4 of\citep{Ma2018}). Very recently, \citet{Shibata2019} reestimated $M_{\rm TOV}\in [2.1,~2.25]M_\odot$ with the data of GW170817 and the main improvements are the calculations of the angular momentum ``carried away" via the neutrino emission, the gravitational wave radiation, the mass ejection and the torus. These authors adopted a group of representative EoSs and then investigated the amount of angular momentum and energy losses needed to trigger the collapse of the remnant formed in the merger.

In \citet{Ma2018}, one work from this group, the constraint on $M_{\rm max}$ has not taken into account the contribution of the torus (this is also the case for some relevant literature) and the angular momentum of the precollapse remnant has simply been assumed to be in a wide range. In this work, we incorporate the approaches of \citet{Shibata2019} on various angular momentum losses, together with the recent improved understanding of the torus and the central engine of GW170817/GRB170817A/AT2017gfo (e.g., see\citep{WangYZ2019,Gill2019}, Sec.3 for the details), to estimate dimensionless angular momentum of the remnant and then $M_{\rm TOV}$. For our purpose, instead of carrying out individual EoS case studies, we adopt some EoS-insensitive relationships, which have been suggested in the literature but updated in this manuscript, to analytically derive an approximated expression for the critical total gravitational mass ($M_{\rm tot,c}$) to form SMNS in the DNS mergers.
With some information from the observations and numerical simulations, we finally have a $M_{\rm TOV}=2.13^{+0.09}_{-0.08}M_\odot$ (68.3\% credibility interval), which is consistent with the constraints set by current NS mass measurements.

\section{The critical total gravitational mass for forming SMNS in DNS merger}
\subsection{The general formulas}

As depicted in \citet{Hanauske2017} (see also \citep{Fujibayashi2018}), in the very early phase of DNS merger, the remnant is a rapidly rotating core surrounded by an outer torus rotating in nearly Keplerian state and attached to it continuously. The evolution of the remnant depends on the total mass and rotating configuration of the system, as well as EoS. The fate of the core of the remnant will be a promptly formed black hole, a stable neutron star, or a transient HMNS/SMNS which will finally collapse into a black hole.

By definition, the baryonic mass ($M_{\rm b}$) of an NS consists of the binding energy ($BE$) and the gravitational mass ($M$) of an equilibrium configuration. In the scenario of a DNS merger, the conservation of the baryonic mass before and after coalescence yields
\begin{equation}
M_1+M_2+BE_1+BE_2=M_{\rm b,rem}+m_{\rm loss},
\label{eq:Mbr}
\end{equation}
where the subscripts $1,2$, and ${\rm rem}$ are for progenitors and the core of remnant, respectively, and $m_{\rm loss}$ is all the mass outside of the core, including the masses of surrounding torus and the ejecta.

\citet{Lattimer2001} found first that binding energy is essentially independent of the EoS and gave out an expression of $BE$ with respect to the compactness of the neutron star, i.e., $BE/M=0.6\zeta/(1-0.5\zeta)$, where the compactness $\zeta=GM/Rc^2$, $R$ is the radius of NS and $c$ is the speed of light. \citet{Breu2016} extended this work to more modern EoSs and yields $BE/M=0.619\zeta+0.136\zeta^2$.

Out of our numerical calculations with the RNS code \citep{1995ApJ...444..306S, 1992ApJ...398..203C, 1989MNRAS.237..355K} and a set of EoSs satisfying $M_{\rm TOV}\geq 2.0 M_\odot$ and $R_{1.4}<14$ km (please see Appendix for details, where the subscript $1.4$ represents the NS mass of $1.4M_\odot$), we found a new expression of BE and $\zeta$ (for $0.05\leq \zeta \leq 0.25$ since we are interested in the neutron stars in the DNS systems), which reads
\begin{equation}
\frac{BE}{M}=-0.0130+0.618\zeta+0.267\zeta^2.
\label{eq:BE1}
\end{equation}

In the case of the rotating neutron star, there is always a turning point for a neutron star with a certain angular momentum from slow rotating up to the Keplerian limit, where the neutron star resists more gravity than in the state of nonrotating and stays in equilibrium with maximum gravitational mass ($M_{\rm crit}$) under this angular momentum.
Our calculations revealed a universal relation among $M_{\rm crit},~j$, and $\zeta_{\rm TOV}$, i.e.,
\begin{equation}
\frac{M_{\rm crit}}{M_{\rm TOV}}=1+9.02\times10^{-2}\zeta_{\rm TOV}^{-1}j^2+1.93\times10^{-2}\zeta_{\rm TOV}^{-2}j^4,
\label{eq:Mcr}
\end{equation}
where $j\equiv cJ/GM^2$ is the dimensionless angular momentum and $\zeta_{\rm TOV}=GM_{\rm TOV}/R_{\rm TOV}c^2$ is the compactness of the neutron star in the maximum configuration of nonrotating state. Although the turning points we found are similar to those found by \citet{Breu2016}, our relation differs from their Eq.(18) because these authors simply took $I/MR^2$ as constant, while in reality it increases almost linearly with $\zeta_{\rm TOV}$ (please see Appendix for the extended discussion).
The binding energy at this point follows the empirical relation of
\begin{equation}
\begin{aligned}
\frac{BE_{\rm crit}}{M_{\rm crit}}=-0.10+0.78(1-0.050j-0.034j^2)\zeta_{\rm TOV}+0.61(1+0.23j-0.58j^2)\zeta_{\rm TOV}^2.
\label{eq:BEr}
\end{aligned}
\end{equation}

With Eqs.(\ref{eq:Mcr}) and (\ref{eq:BEr}), we have
\begin{eqnarray}
M_{\rm b,crit}&=& M_{\rm crit}+BE_{\rm crit}\nonumber\\
&=& M_{\rm TOV} K_j(j, \zeta_{\rm TOV})\nonumber\\
&=& M_{\rm TOV} (1+9.02\times10^{-2}\zeta_{\rm TOV}^{-1}j^2+1.93\times10^{-2}\zeta_{\rm TOV}^{-2}j^4)\nonumber\\
&&\times [0.90+0.78(1-0.050j-0.034j^2)\zeta_{\rm TOV}+0.61(1+0.23j-0.58j^2)\zeta_{\rm TOV}^2]
\label{eq:Mbcrit}
\end{eqnarray}

For given $M_{\rm TOV}$ and $R_{\rm TOV}$, $M_{\rm b,crit}$ depends on the dimensionless angular momentum parameter $j$.
With Eqs.(\ref{eq:Mbr})$-$(\ref{eq:Mbcrit}) and setting $M_{\rm b,rem}=M_{\rm b,crit}(j)$, we obtain a sequence of critical values of total gravitational mass of DNS, $M_{\rm tot,c}(j)$, which determines the fate of the remnant of the DNS merger.
For $M_{\rm tot} \le M_{\rm tot,c}(j=0)$, the remnant core will be a stable neutron star. While $M_{\rm tot} > M_{\rm tot,c}(j=j_{\rm kep})$, it will form a transient HMNS or collapse to a black hole promptly, where $j_{\rm kep}$ represents the dimensionless angular momentum at the Keplerian limit, which is $\sim 0.7$ for most EoS models. For $M_{\rm tot,c}(j=j_{\rm kep}) \ge M_{\rm tot} > M_{\rm tot,c}(j=0)$, it will become a transient SMNS, and finally collapse to a black hole when its angular momentum has been dissipated to a value $j_{\rm coll}$ that satisfies $M_{\rm tot} = M_{\rm tot,c}(j=j_{\rm coll})$.

As shown above, $M_{\rm tot,c}$ is sensitive to the general parameters of ($M_{\rm TOV},~R_{\rm TOV}$) and the parameters of ($j,~m_{\rm loss}$) which are needed to be evaluated case by case. Below we will demonstrate that the dependences on other parameters are rather insensitive.

In the following analytic/simplified approach, we express the mass in units of solar mass ($M_\odot$). Benefiting from the fact that $\zeta$ is relatively small and narrowly distributed, $BE\approx (-0.020+0.71\zeta) M$ is found to be a good approximation. Assuming a DNS system with total mass $M_{\rm tot}$, mass difference $\Delta\equiv (M_1-M_2)/M_{\rm tot}$ and radius difference $\delta \equiv (R_2-R_1)/R_2$ (hereafter $R=R_1$) between two components that coalesce and collapse to a black hole after a transient phase of SMNS, then Eq.(\ref{eq:Mbr}) becomes

\begin{equation}
0.980M_{\rm tot,c}+0.0437M_\odot^{-1}(1+\Delta^2+\Delta\delta-\delta/2-\delta\Delta^2/2)(R/12~{\rm km})^{-1}M_{\rm tot,c}^2=M_{\rm TOV}K_j(j,\zeta_{\rm TOV})+m_{\rm loss},
\label{eq:Mbr-2}
\end{equation}
where we have normalized the radius of a typical NS (with a mass of $\sim 1.4M_\odot$) to $12$ km, a value favored by the data of GW170817 and other astrophysical constraints (e.g.,\citep{Abbott2018,Jiang2019}).
The current EoS models usually predict a very small $\delta$ for the masses of the neutron stars accurately measured in the Galactic systems (e.g. in the range between 1.1 and $1.5M_\odot$ it is about 0.002 for model WFF2, 0.007 for APR4, and $\sim 0.015$ for the best fitted mass-radius relation of \citet{Ozel2016b}), while 14 out of all 17 of those systems (see\citep{Huang2018} and the references therein) have mass difference less than $0.4~M_\odot$ (i.e. $\Delta\leq0.15$). So we ignore the item of $\delta\Delta^2$, and the solution of Eq.(\ref{eq:Mbr-2}) turns out to be
\begin{equation}
M_{\rm tot,c}\approx 11.2M_\odot\frac{(R/12{\rm km})}{(1+\Delta^2+\Delta\delta-\delta/2)}\left\{-1+\sqrt{1+0.182\left[\frac{(1+\Delta^2+\Delta\delta-\delta/2)}{(R/12{\rm km})}\right]\left(\frac{M_{\rm TOV}K_j+m_{\rm loss}}{1M_\odot}\right)}\right\}.
\label{eq:main-1}
\end{equation}

\begin{figure}
\centering
\includegraphics[width=4in]{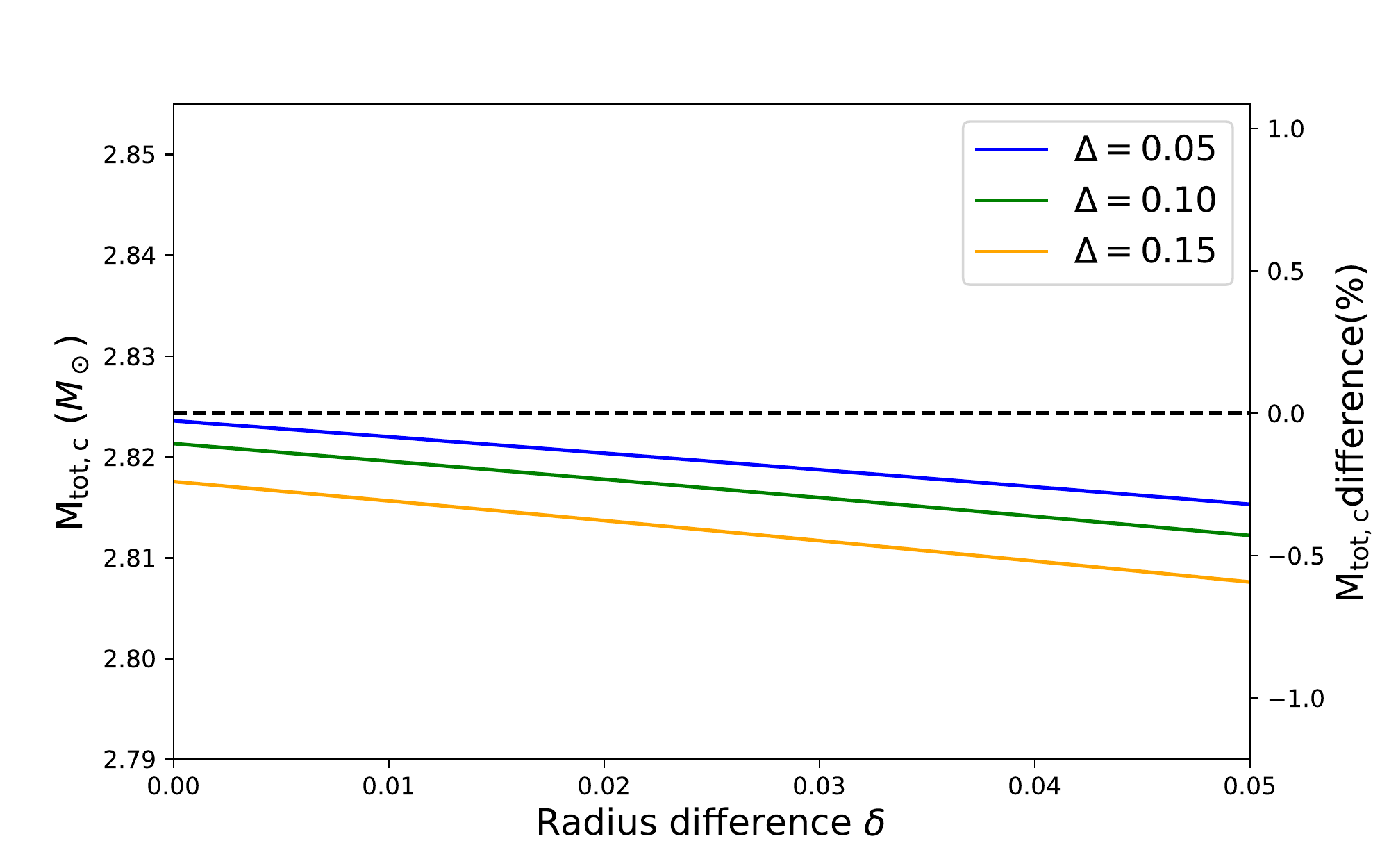}
\caption{$M_{\rm tot,c}$ versus $\Delta$ and $\delta$. The physical parameters are adopted as $R=12~{\rm km}$, $R_{\rm TOV}=10~{\rm km}$, $M_{\rm TOV}=2.14~M_\odot$, $m_{\rm loss}=0.1~M_\odot$, and $j=0.7$. The dashed line is the fiducial value for the symmetric system.}
\label{fig:delta}
\hfill
\end{figure}

Compared to the symmetric scenario (the progenitors are identical), the differences of mass and radius of progenitors only make the calculation of $M_{\rm tot,c}$ slightly deviated(see the numerical example presented in Fig.\ref{fig:delta}). With the parameters of $\Delta\in [0,0.15]$ and $\delta \in [0,0.02]$, the induced correction to $M_{\rm tot,c}$ is less than 1\% and can be safely ignored.

As shown in Appendix [i.e., Fig.\ref{fig:fit_BEcrit}], $BE_{\rm crit}/M_{\rm crit}$ is insensitive to $j$, and then we further approximate $K_j$ as
\begin{equation}
\bar{K}_j\approx (1+9.02\times10^{-2}\zeta_{\rm TOV}^{-1}j^2+1.93\times10^{-2}\zeta_{\rm TOV}^{-2}j^4)(0.8634+1.051\zeta_{\rm TOV}).
\label{eq:Kj-1}
\end{equation}

With Eqs.(\ref{eq:main-1}) and (\ref{eq:Kj-1}), it is straightforward to yield the relation between $M_{\rm tot,c}$ and $M_{\rm TOV}$, i.e.,
\begin{eqnarray}
M_{\rm tot,c}
&\approx 11.2M_\odot~(R/12~{\rm km}) \left[-1+\sqrt{1+0.182M_\odot^{-1}\left(M_{\rm TOV}\bar{K}_j+m_{\rm loss}\right)(R/12{\rm km})^{-1}}\right]\nonumber\\
&\approx 1.05M_{\rm TOV}\bar{K}_j\left[1-0.0455M_\odot^{-1}\left(M_{\rm TOV}\bar{K}_j+2~m_{\rm loss}\right)\left(R/12{\rm km}\right)^{-1}\right]+m_{\rm loss}.
\label{eq:main-2}
\end{eqnarray}
Equations(\ref{eq:Kj-1})$-$(\ref{eq:main-2}) show that $M_{\rm tot,c}$ depends very insensitively on $R$ but relatively sensitive on $M_{\rm TOV}$, $j$, and $R_{\rm TOV}$ (or alternatively $\zeta_{\rm TOV}$).
The second term of Eq.(\ref{eq:main-2}) is about 12\% of the first, with such an approximation we further simplify Eq.(\ref{eq:main-2}) into
\begin{equation}
M_{\rm tot,c}\approx 0.924M_{\rm TOV}(1+7.94\times10^{-2}\zeta_{\rm TOV}^{-1}j^2+1.70\times10^{-2}\zeta_{\rm TOV}^{-2}j^4)(0.8634+1.051\zeta_{\rm TOV})(1-0.091~M_\odot^{-1}~m_{\rm loss})+m_{\rm loss}.
\label{eq:main-3}
\end{equation}

\begin{figure}
\centering
\subfigure{\label{fig:MmaxJ}
\includegraphics[width=3.3in]{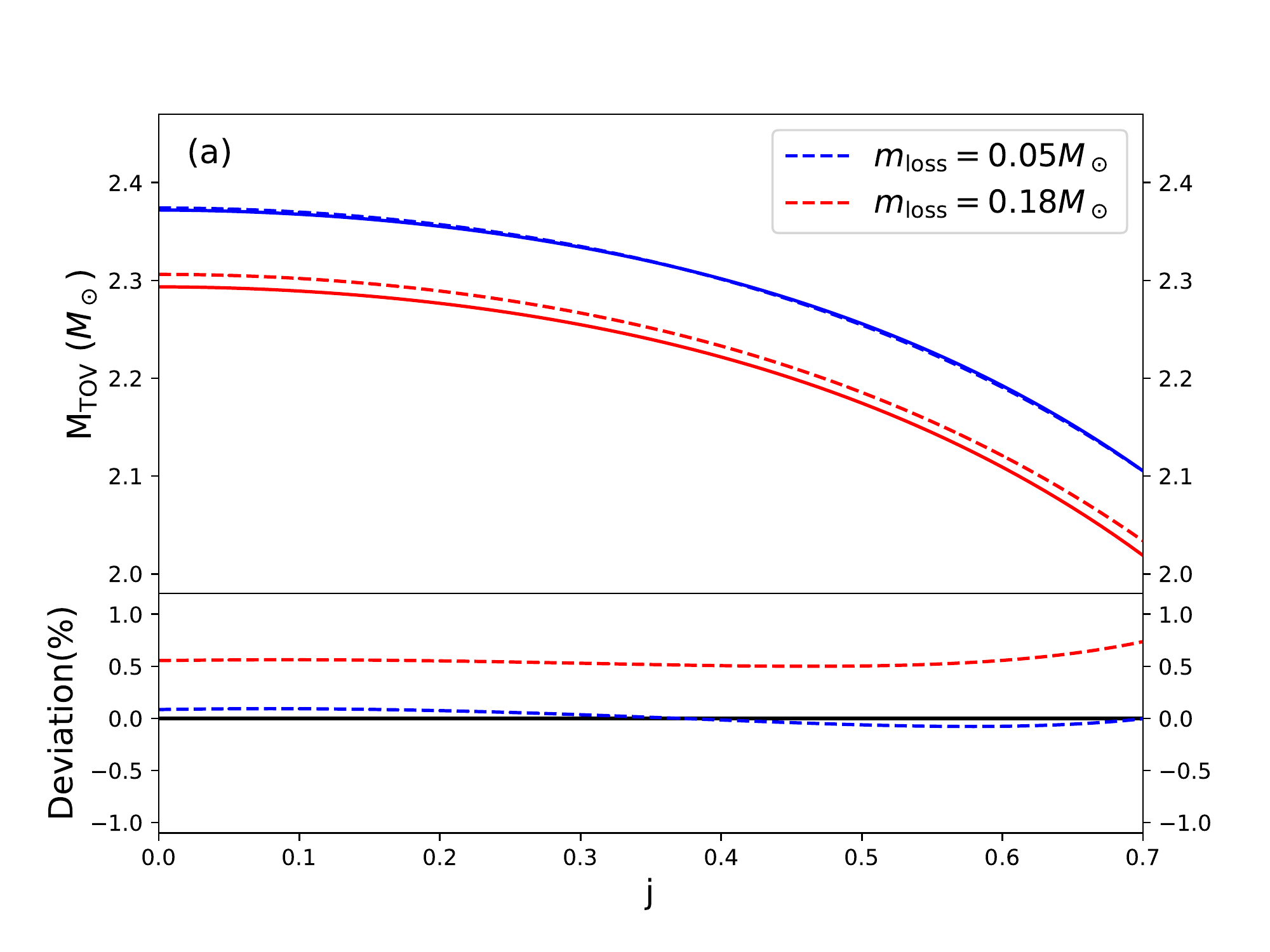}}
\subfigure{\label{fig:Mmaxmloss}
\includegraphics[width=3.3in]{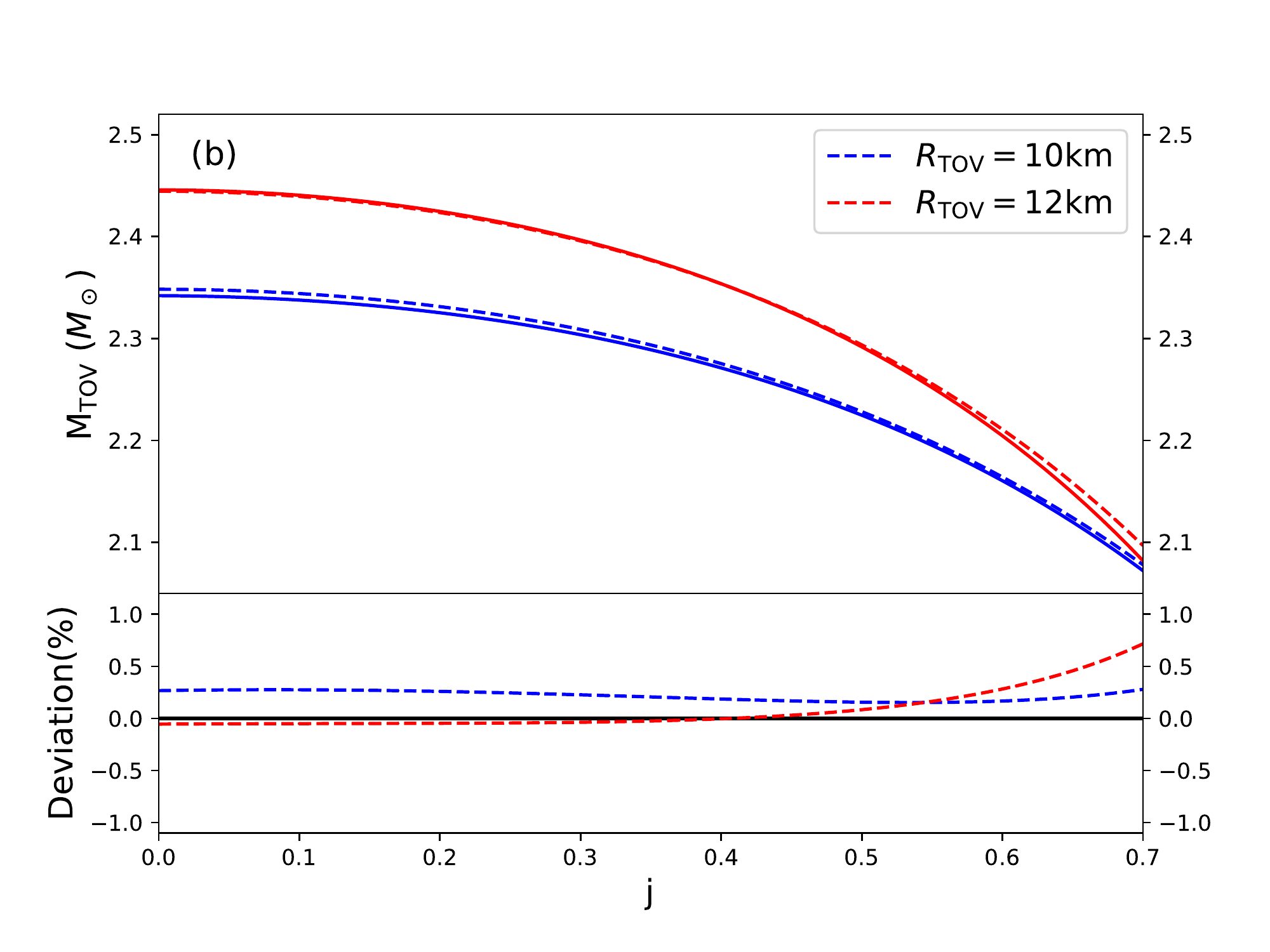}}
\subfigure{\label{fig:MmaxJ-R}
\includegraphics[width=3.3in]{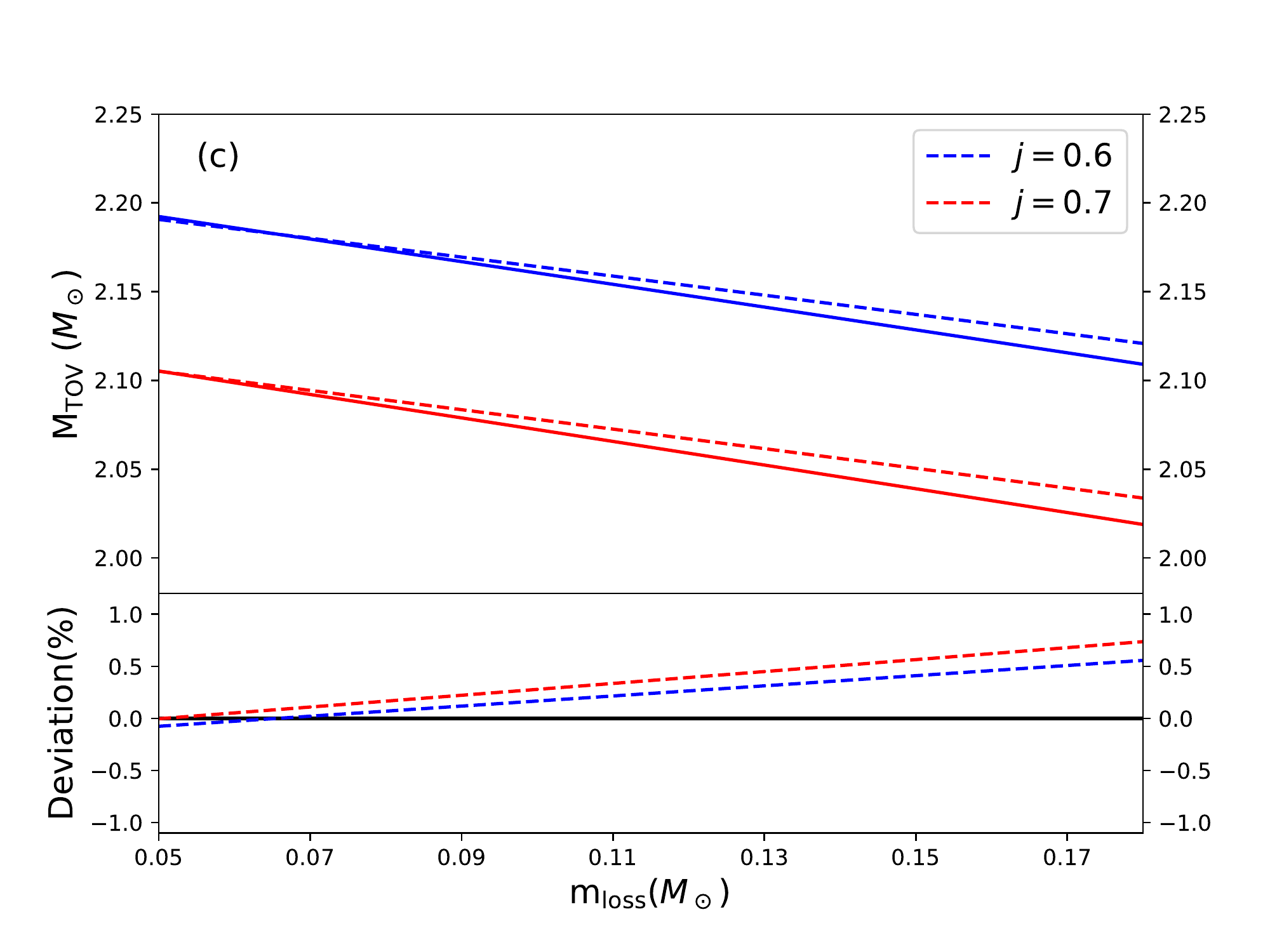}}
\caption{(a) $M_{\rm TOV}$ vs $j$ with $m_{\rm loss}=(0.05,0.18)~M_\odot$, and (c) $M_{\rm TOV}$ vs. $m_{\rm loss}$ for $j_{\rm coll}=(0.6, 0.7)$, supposing that $M_{\rm tot,c}=2.74M_\odot$, $R=12$ km and $R_{\rm TOV}=10$ km. (b) $M_{\rm TOV}$ vs $j$ with $R_{\rm TOV}=(10,~12)~\rm km$, while $M_{\rm tot,c}=2.74M_\odot$ and $m_{\rm loss}=0.1M_\odot$. The solid and dashed lines are the solutions of Eq.(\ref{eq:Mbr})-(\ref{eq:Mbcrit}) and the universal relation Eq.(\ref{eq:main-3}), respectively.}
\hfill
\label{fig:Mmax-J}
\end{figure}

As demonstrated in Fig.\ref{fig:Mmax-J}, this approximated/simplified expression is well consistent with the numerical results obtained through eqs.(\ref{eq:Mbr})$-$(\ref{eq:Mbcrit}).
Therefore, for a DNS merger event with plentiful electromagnetic counterpart data, once there is the evidence for the formation/collapse of a transient SMNS and both $m_{\rm loss}$ and $j_{\rm coll}$ (the angular momentum of the core of the remnant at the onset of collapse) have been reasonably estimated,
we can deduce the TOV properties of neutron star via the relation of $M_{\rm tot,c}(j_{\rm coll})=M_{\rm tot}$.

In the precollapse phase, if the kinetic rotational energy of the newly formed remnants carried away by the viscosity generated by the magnetohydrodynamic turbulence, gravitational wave radiation and the neutrinos is efficient (e.g.,\citep{Hotokezaka2013a,Bernuzzi2016}), we would expect $j<j_{\rm kep}(\approx 0.7)$; otherwise, we have $j\approx j_{\rm kep}$. Unless the gravitational wave can carry away most of the kinetic energy of the remnant, it is less likely to have $j \leq 0.5$ (see the discussion in the next subsection), and therefore we simply take two representative values of $j=(0.6,~0.7)$, respectively. Usually $m_{\rm loss}$ is dominated by the masses of the subrelativistic ejecta and the surrounding torus.

\subsection{The angular momentum of the precollapse remnant}

For a giving set of $(M_{\rm TOV},~R_{\rm TOV},~M_{\rm tot},~j,~m_{\rm loss})$, it is straightforward to estimate the fate of the remnant formed in the binary neutron star merger. However, estimating $j$ and $m_{\rm loss}$ are not easy tasks, which are the focus of the following discussion.
In this work we concentrate on the scenario that a HMNS and subsequently a SMNS formed right after the DNS merger, which collapses to a black hole at the time of $t_{\rm coll}$ (the merger time is taken to be the zero point). The value of $m_{\rm loss}$ should be the sum of $M_{\rm torus}$ and $M_{\rm eje}$ for $t\leq t_{\rm coll}$. Here $M_{\rm eje}$ contains the whole dynamical ejecta ($M_{\rm eje,dyn}$) launched promptly, and parts of late-time (before collapse) neutrino-driven ejecta ($M_{\rm eje,\nu}$) and magnetically driven ejecta($M_{\rm eje,B}$) from the surrounding torus.

To estimate the value of dimensionless angular momentum ($j_{\rm coll}$) at the onset of collapse, we need to consider angular momentum conservation of the system carefully.
The angular momentum of the progenitors at the onset of merger ($J_0$) must conserve with the angular momenta of the remnant core itself {\bf ($J_{\rm coll}$)} and the parts outside ($J_{\rm eje}+J_{\rm torus}$), as well as those carried away by radiations of postmerger gravitational wave and neutrinos ($J_{\rm GW,p}+J_{\rm \nu}$). The angular momenta of these components, depending on their properties such as mass, radius, energy ($E_{\rm GW,p}$) and frequency ($f$) of GW, energy ($E_{\rm \nu}$) and location of neutrino radiation, can be obtained semianalytically together with appropriate approximations with the help of numerical relativity simulations. Recently, this problem has been systematically examined by
\citet{Shibata2019}, and in this work we follow their approaches with some modifications/discussions. Based on the numerical simulations, these authors suggested an empirical relation among $J_0$, $M_{\rm tot}$, the symmetric mass ratio ($\eta$), and radius of neutron star with a gravitational mass of 1.35 $M_\odot$ ($R_{\rm 1.35}$), which reads
\begin{equation}
J_0\approx G c^{-1} M_{\rm tot}^2 \eta \left[a_1 - a_2 \delta\eta + a_3 \left(\frac {R_{\rm 1.35}}{\rm 10~km}\right)^3 \left(1+a_4 \delta\eta \right)\right],
\label{eq:J_0}
\end{equation}
where $\delta\eta=\eta-1/4$, $a_1\approx3.32$, $a_2\approx31$, $a_3\approx 0.137$, and $a_4\approx27$.
The angular momentum of the torus and the ejecta are estimated as
$J_{\rm torus}\approx M_{\rm torus}\sqrt{G M_{\rm rem}R_{\rm torus}}$
and
$J_{\rm eje}\approx M_{\rm eje}\sqrt{G M_{\rm rem}R_{\rm eje}}$,
respectively, where $R_{\rm torus}$ is the typical radius of torus, $R_{\rm eje}$ is the typical distance from merger center to where ejecta occur and $M_{\rm rem}$ is the gravitational mass of remnant core.
For dynamical ejecta launched within a dynamical timescale ($\leq 10~{\rm ms}$), $R_{\rm eje}$ can be estimated as the typical radius of the surrounding torus at such an early time. The late-time neutrino-driven and magnetically driven ejecta mainly come from surrounding torus with $R_{\rm eje}=R_{\rm torus}$ (see also\citep{Shibata2019}) while others come from the core of the remnant with $R_{\rm eje}=R_{\rm rem}$.

Precollapse, a large number of neutrinos radiate from the hot, violent, and rapidly rotating remnant core and make it cool down and spin down. \citet{Baumgarte1998} estimated the angular momentum carried away by neutrinos as
$J_{\rm \nu}\approx (2/3)c^{-2}R_{\rm rem}^2\Omega E_\nu \approx2.2\times10^{48}\rm erg~s(E_{\rm \nu}/{0.1M_\odot c^2})(R_{\rm rem}/{13\rm km})^{2}
(\Omega/{10^4\rm rad/s})$,
where $\Omega$ is the angular velocity of the rigidly rotating massive neutron star. The equatorial radius $R_{\rm rem}$ can be approximated as $R_{\rm crit}=(1+0.032\zeta_{\rm TOV}^{-1.6}j^2+0.014\zeta_{\rm TOV}^{-3.2}j^4)R_{\rm TOV}$ (see Appendix for details). For a neutron star rotating in the Keplerian limit with $R_{\rm TOV}$ = 10 km, we have $R_{\rm rem}\approx 13$ km.

The angular momentum carried away by a postmerger gravitational wave can be expressed as 
$J_{\rm GW,p}= J_0(1-\sqrt{1-E_{\rm GW,p}/E_0})$,
 where the rotating energy at the onset of the merger can be approximated as
$E_0 = \frac{1}{2}I\Omega^2 \approx 1.5\times10^{53}\rm erg\left(\rm I /{3\times10^{45} g~cm^2}\right)\left({\Omega}/{10^4~rad/s}\right)^2$.
Since the frequency of the gravitational wave ($f$) is double of the spin frequency of the remnant core, for $E_{\rm GW,p} \ll E_0$ we obtain an approximation as
$J_{\rm GW,p} \approx E_{\rm GW,p}/\pi f
\approx9.5\times10^{48}\rm erg~s\left(E_{\rm GW,p}/{0.05M_\odot c^2}\right)\left(f/{3.0\rm kHz}\right)^{-1}$,
which is consistent with that of \citet{Shibata2019}.
While for $E_{\rm GW,p} \approx E_0$, we have $J_{\rm GW,p}\approx J_0$, implying that the most angular momentum has been carried away by the gravitational wave radiation.

Finally, we have
\begin{equation}
\begin{split}
J_{\rm coll} &\approx J_0 - J_{\rm eje} - J_{\rm torus} - J_{\rm GW,p} - J_{\rm \nu}, \\
j_{\rm coll}&=cJ_{\rm coll}/GM_{\rm rem}^2.
\end{split}
\label{eq:j_rem}
\end{equation}

\section{The estimate of $M_{\rm TOV}$ with GW170817} \label{Sec:MTOV-GW170817}
As outlined above, with Eq.(\ref{eq:main-3}) we can constrain the maximum mass of nonrotating neutron star by the observations of the DNS merger events, especially if the electromagnetic counterparts (ideally, both the GRB/afterglow and the kilonova signals) are available. So far, the LIGO/Virgo detectors have detected the DNS merger-driven gravitational wave signals from GW170817 \citep{Abbott2017a}, GW190425 \citep{Abbott2020} (note that for this event the neutron star$-$black hole merger scenario is still possible \citep{Han2020}), and possibly also S190901ap, S190910h, S191213g and S200213t.\footnote{\url{https://gracedb.ligo.org/superevents/public/O3/}} No reliable counterparts for GW190425, S190901ap, S190910h, S191213g and S200213t have been identified. While for GW170817, the associated $\gamma$-ray burst GRB 170817A, though weak, had been robustly detected by the {\it Fermi} Gamma-ray Monitor before the gravitational wave alert \citep{Goldstein2017}. The follow-up multiwavelength observations successfully detected/identified the kilonova emission as well as the off-axis afterglow emission (see, e.g., \citep{Abbott2017b,Lamb2018,Duan2019} and the references therein).

The analysis of the gravitational wave data of GW170817 yields a total gravitational mass $M_{\rm tot}=2.74^{+0.04}_{-0.01}~M_\odot$ of the premerger neutron stars. Although the fate of the remnant is still not very clear, the most widely accepted hypothesis is a short-lived massive neutron star that had collapsed to a black hole within $\sim 1$ s. Recently \citet{Gill2019} combined the constraints from the blue kilonova and the time lag between GW and GRB signals and finally inferred a collapse time of $t_{\rm coll}=0.98^{+0.31}_{-0.26}~{\rm s}$ for the massive remnant formed in the merger event GW170817. Since the initial differential rotation of the massive remnant may have been terminated by the magnetic braking within $\sim 0.1~{\rm s}$ (e.g., \citep{Hotokezaka2013b,Shibata2017}; see, however, \citep{Ruiz2018}), it might be reasonable to assume that the precollapse object was an SMNS. In this work {\it we adopt such an assumption for GW170817} [i.e., at $t_{\rm coll}\sim 1$ s the SMNS, with a dimensionless angular momentum $j_{\rm coll}$, collapsed to a BH; for which we can replace the $M_{\rm tot,c}$ in Eq.(\ref{eq:main-3}) by $M_{\rm tot}$ to estimate $M_{\rm TOV}$] and then apply the framework developed above to constrain on $M_{\rm TOV}$. For such a purpose we need to know both $j_{\rm coll}$ and $m_{\rm loss}$.

\citet{Hanauske2017} numerically simulated the evolution of massive NSs formed in DNS mergers. By a reasonable criteria to divide the core of the remnant and surrounding torus, they found the mass of torus at the onset of collapse is in the range of 0.014$-$0.059$M_{\rm tot}$, which corresponds to 0.04$-$0.16$M_\odot$ in the case of GW170817. Other numerical simulations (see \citep{Shibata2019} and the references therein) also found $\sim$ 0.1$-$0.2$M_\odot$ mass of the surrounding torus at the early stage and decreases to $\sim 0.02-0.05M_\odot$ at $t\sim 1~{\rm s}$. For the remnant neutron star that collapsed very quickly (i.e., at the early stage), about half of the torus material in the vicinity will be swallowed by the black hole at its formation. However, for $t_{\rm coll}\sim 1~{\rm s}$, such a sudden torus mass drop is unlikely. The reason is the following. The inner part of the torus material has transported their angular momentum to the outer material via the viscosity process. As a result, the inner torus materials have become part of the central remnant and the outer torus material expands outward. The corresponding radius of the late time torus is $R_{\rm torus}\approx 40+100(t_{\rm coll}/1{\rm s})^{1/2}~{\rm km}$ \citep{Fujibayashi2018}, which is significantly larger than the radius of the innermost stable circular orbit of the ``nascent" Kerr black hole that is expected to be $\sim 15~{\rm km}$. Therefore, we do not expect significant mass accretion at the collapse time of the SMNS, and the torus mass can be reasonably approximated by the accretion disk mass inferred in the modeling of GRB 170817A and its afterglow \citep{WangYZ2019}.
GRB 170817A is most likely from an off-axis structured jet \citep[see][for the theoretical argument]{Jin2018} and thus very weak. Fortunately, the plentiful x-ray, optical, and radio afterglow data are very helpful in revealing the properties of the GRB ejecta \citep[e.g.][]{Lamb2018} and hence the mass of the accretion disk, which is found to be $M_{\rm torus}\sim$ 0.015$-$0.134$M_\odot$ (90\% confidence level; and the most plausible value is $0.035M_\odot$), as shown in \citet{WangYZ2019}. Interestingly, such a mass range is well consistent with that found/suggested in the numerical simulations (note that in the estimate of \citet{Shibata2019} a torus mass of 0.02$-$0.1$M_\odot$ at the collapse time was adopted).
As a cross-check, we have also directly taken into account the mass range suggested by  Hanauske \etal (\citep{Hanauske2017},assuming a flat probability distribution) in our calculation and obtained quite consistent results.

Following \citet{Gill2019} we integrate a mass ejection rate within $t_{\rm coll}$ and find $M_{\rm eje}\sim$0.03$-$0.04$M_\odot$, which is well consistent with that found in the kilonova modeling (e.g.,\citep{Pian2017,Kawaguchi2018}).
The estimate of $j_{\rm coll}$, however, is more challenging because there are no direct measurements for some key information and we have to make some approximations based on recent numerical simulations
(e.g.,\citep{Baumgarte1998,Bauswein2012,Kastaun2013,Hotokezaka2013b,Radice2016,Sekiguchi2016,Ciolfi2017,Shibata2017,Radice2018,Fujibayashi2018,Shibata2019}). For the merger of a binary neutron star with a total mass $M_{\rm tot}=2.74M_\odot$ and a mass ratio of $q\sim$ 0.73$-$1.00, the numerical simulations yield a $J_0\sim 5.8-6.3\times10^{49}~{\rm erg~s}$ \citep{Shibata2019}.
The dominant gravitational waves emitted in the postmerger phase were found to be with the frequency $f\in[2,4]$ kHz \citep{Bauswein2012,Hotokezaka2013b}. With Eq.(5) of \citet{Zappa2018} and the posterior distribution of the tidal deformabilities of the two neutron stars involved in GW170817, the peak gravitational wave luminosity is estimated to be $\sim 2\times 10^{55}~{\rm erg~s^{-1}}$. The postmerger gravitational wave radiation can emit in total about 0.8\%$-$2.5\% of mass energy of the DNS system \citep{Bernuzzi2016}. Therefore, for GW170817 we take $E_{\rm GW,p} \subset(0.02,~0.07) M_\odot c^2$.

The neutrino radiation happened right after the merger when the violate collision made the remnant extremely hot. The current numerical simulations suggest a luminosity of $L_{\nu} \sim 2\times10^{53}~{\rm erg~s^{-1}}$ \citep{Baumgarte1998,Sekiguchi2016,Foucart2016,Radice2016}, and then the energy of neutrino radiation can be estimated as $E_{\nu}=L_\nu t_{\rm coll}$.

Based on the analysis of the gravitational wave data of GW170817, we assume $R_{\rm TOV}\in[0.8,0.95]R_{1.4}$ with a Beta-PERT distribution peaking at 0.90 (an additional request is that $R_{\rm TOV}>9.6{\rm km}$, as shown in\citep{Bauswein2017}).
Finally, we assume all the parameters uniformly distributed, except $M_{\rm tot}$, $R_{\rm 1.4}$, $\eta$ and $M_{\rm torus}$, for which the distributions are adopted from \citet{Abbott2017a} and \citet{WangYZ2019}, respectively.

\begin{figure}
\centering
\includegraphics[width=3.3in]{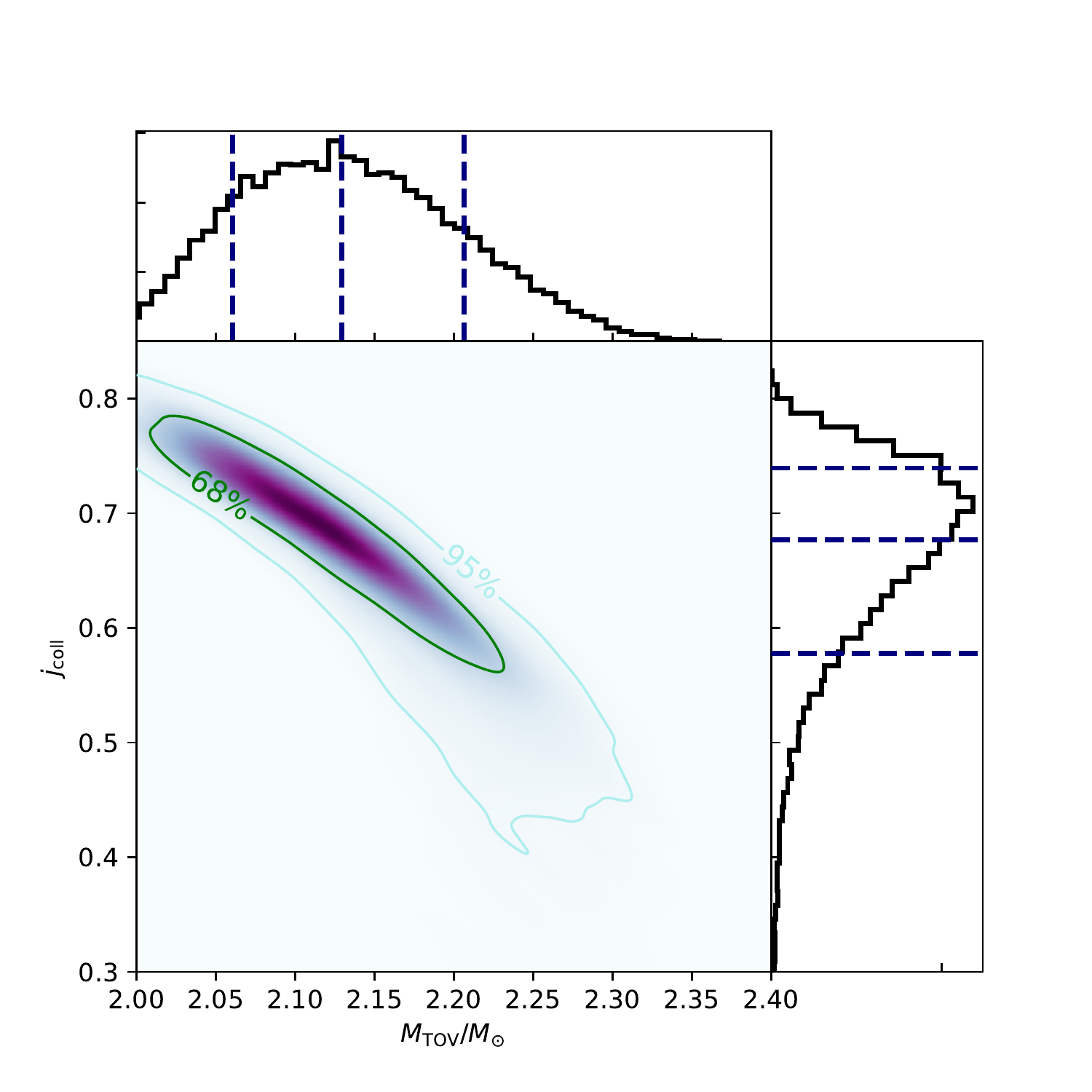}
\caption{The probability density of $j_{\rm coll}$ and $M_{\rm TOV}$ in the case of GW170817/GRB 170817A/AT2017gfo. Please see the main text for the parameters adopted in the analysis. The underlying assumption is that the central engine launching the GRB 170817A was a black hole quickly formed from the collapse of a transient SMNS.}
\label{fig:final}
\hfill
\end{figure}

Then we generate $3\times10^{4}$ groups of samples with Monte Carlo methods using Eqs.(\ref{eq:main-3}), (\ref{eq:J_0}), and (\ref{eq:j_rem}) and the distributions mentioned above [please note that now we replace $M_{\rm tot,c}$ in Eq.(\ref{eq:main-3}) by $M_{\rm tot}$, as explained in the second paragraph of this section]. Thus, the baryonic mass and angular momentum conservation equations are closed, with which we can solve the unknown parameters, i.e., $j_{\rm coll}$ and $M_{\rm TOV}$, for each group of samples. Combining all these solutions, we further obtain the probability density distributions of $j_{\rm coll}$ and $M_{\rm TOV}$, as shown in Fig.\ref{fig:final}. The $68.3\%$ confidential interval of $M_{\rm TOV}$ is $2.13^{+0.08}_{-0.07}M_\odot$ (for the $95\%$ confidential interval, we have $M_{\rm TOV}=2.13^{+0.17}_{-0.11}M_\odot$) and $j_{\rm coll}$ is close to the mass shedding limit in most cases. Our result is not exactly the same as, though is close to, that of \citet{Shibata2019} because of the differences in approaches and some assumptions. Our result is also consistent with the upper limits set in the literature with the data of GW170817/GRB 170817A/AT2017gfo (e.g., \citep{Margalit2017,Shibata2017,Rezzolla2018,Ma2018,Ruiz2018}).

In the above approach, the systematic uncertainties of the adopted empirical relationships (see Fig.\ref{fig:fit_Mcrit}, \ref{fig:fit_BE} and \ref{fig:deviation_fit_Mcrit} in Appendix) have been ignored. With the parameters of $\overline {BE}_{\rm crit}\equiv BE_{\rm crit}/M_{\rm crit}$ and $\overline M\equiv M_{\rm crit}/M_{\rm TOV}$ defined in the Appendix, we have
\begin{equation}
M_{\rm TOV}=(M_{\rm tot}+BE_1+BE_2-m_{\rm loss})/[\overline M (1+\overline {BE}_{\rm crit})].
\end{equation}
Therefore the uncertainty of $M_{\rm TOV}$ directly transferred from the empirical relationships can be estimated as
\begin{equation}
\delta M_{\rm TOV,f}=\sqrt{\frac{(\delta BE_1)^{2}+(\delta BE_2)^{2}}{{\overline M}^2(1+\overline {BE}_{\rm crit})^2}+M_{\rm TOV}^2\left[\left(\frac{\delta \overline M}{\overline M}\right)^2+\frac{(\delta \overline {BE}_{\rm crit})^2}{(1+\overline {BE}_{\rm crit})^2}\right]}.
\label{eq:syserror}
\end{equation}
As shown in Fig.\ref{fig:fit_BE}, the largest deviation of $\overline{BE}$ from its best fit value is by a factor of $7.5\%$ for $BE\approx 0.17M$, which suggests that $\delta BE_1\approx \delta BE_2 \leq 0.018M_\odot$ for $M\approx 1.4M_\odot$. The highest relative deviations of $\overline{M}$ and $\overline{BE}_{\rm crit}$ from the best fit lines are $0.6\%$ [for $\overline{M}\approx 1.2$, see Fig.\ref{fig:fit_Mcrit}] and $7.2\%$ [for $\overline{BE}_{\rm crit}\approx 0.24$, see Fig.\ref{fig:deviation_fit_Mcrit}], respectively. Therefore we have
$\delta \overline M \leq 0.007$ and $\delta \overline{BE}_{\rm crit} \leq 0.016$. Substituting these numbers into Eq.(\ref{eq:syserror}) we have $\delta M_{\rm TOV,f}\leq 0.04M_\odot$, indicating that this term is small but nonignorable.
Additional uncertainty of $M_{\rm TOV}$ arises from the modification of $j$ by the deviations of the best fit functions from the ``true values," which is found to be $\delta M_{\rm TOV,fj}\leq 0.01M_\odot$. Finally we set a more reliable constraint of $M_{\rm TOV}=2.13^{+0.09}_{-0.08}M_\odot$ ($68.3\%$ confidential interval, where the uncertainty terms have been added in quadrature).

\begin{figure}
\centering
\includegraphics[width=4in]{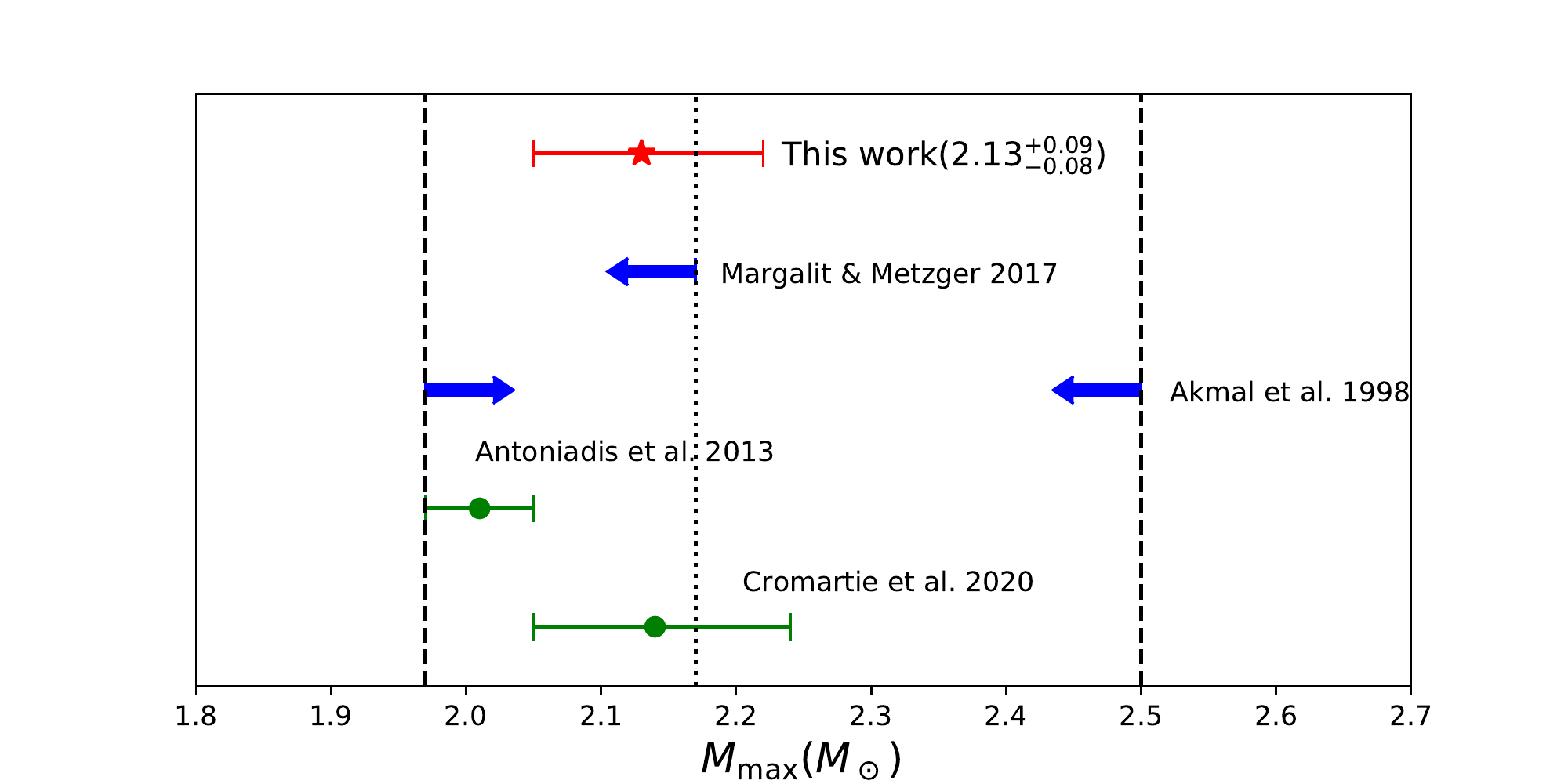}
\caption{Constraints and records of maximum gravitational mass of nonrotating neutron stars. The dashed black lines are the lower and upper limits on $M_{\rm TOV}$, respectively. The former was set by the observation of \citet{Antoniadis2013}, while the latter is from \citet{Akmal1998}. The green dots represent the masses of PSR J0348+0432 and PSR J0740+6620, two accurately measured very-massive pulsars. The dotted black line represents the upper bound induced from GW170817 by \citet{Margalit2017}. Our result $M_{\rm TOV}=2.13^{+0.09}_{-0.08}M_\odot$ is marked in red pentagram. All error bars are for the 68.3\% credibility.}
\label{fig:constraints}
\hfill
\end{figure}

\section{Summary and Discussion} \label{sec:discussion}

Determination of the maximum mass of nonrotating neutron star is of great priority for astrophysics today which will be essential not only to explain a lot of high energy astronomical phenomena but also to understand fundamental nuclei physics. Thanks to the precise measurements of the Shapiro delay in binary MSP systems by pulsar timing, a lower limit of $M_{\rm TOV}>2M_\odot$ has been robustly established \citep{Antoniadis2013,Cromartie2019}. The actual value of $M_{\rm TOV}$, however, is still unknown. In this work we further explore the possibility of estimating $M_{\rm TOV}$ with the double neutron star merger-driven gravitational wave events and their associated electromagnetic counterparts.
We assume that the differential rotation of the massive neutron star formed in DNS mergers has been effectively terminated by the magnetic braking and a uniform rotation has been subsequently established (i.e., a supramassive NS is formed), and we analytically derive Eq.(\ref{eq:main-3}), an approximated relation between $M_{\rm TOV}$ and the critical total gravitational mass ($M_{\rm tot,c}$) to form SMNS in the DNS mergers, benefited from some relationships that are insensitively dependent of equation of state. It shows that $M_{\rm TOV}$, $R_{\rm TOV}$ as well as the dimensionless angular momentum of the core of remnant ($j$) play the dominant roles in modifying $M_{\rm tot,c}$, while the radius and mass differences of the premerger NSs do not. In the case of GW170817, assuming the central engine for GRB 170817A/AT2017gfo was a black hole quickly formed in the collapse of a supramassive NS at $t_{\rm coll}\sim 1~{\rm s}$, we have $M_{\rm TOV}=2.13^{+0.09}_{-0.08}M_\odot$ (68.3\% credibility interval). Such a result is consistent with current bounds set by the neutron star mass measurements as well as some theoretical investigations (see Fig.\ref{fig:constraints}). In reality, other than caused by the loss of angular momentum, the collapse of the merger remnant could also be due to the removal of the thermal pressure support of the thermal neutrinos, for which a smaller $M_{\rm TOV}$ is possible but the recent mass measurement of PSR J0740+6620 is not in support of such a hypothesis.

Though our result is encouraging, there are some uncertainties/cautions in addition to the simplified initial model assumptions. One is that Eq.(\ref{eq:Mbcrit}) is for the cold rapidly rotating massive neutron stars, while the nascent remnants formed in double neutron stars are hot with very strong neutrino radiation. Further studies are needed to check how the estimate of $M_{\rm TOV}$ will be shaped if such an effect has been taken into account. The other is that even for GW170817 with the successful detection of the electromagnetic counterparts (i.e., GRB 170817A/afterglow and AT2017gfo) the $m_{\rm loss}$ is just loosely constrained. In particular, the off-axis nature of GRB 170817A renders the ``intrinsic" radiation energy of the prompt emission quite uncertain (see \citep{WangYZ2019} for the detailed discussion). Because of the lack of the x-ray observations in the first half day after the merger, it is also unclear whether there were x-ray flares that were powered by the reactivity, likely triggered by the fallback accretion of the material ejected during the merger, of the central engine. If yes, these materials might carry a large amount of angular momentum, which enhances $m_{\rm loss}$, reduces $j$ and subsequently modifies $M_{\rm TOV}$ in Sec.\ref{Sec:MTOV-GW170817}. Last but not least, the gravitational wave energy radiated in the precollapse phase of GW170817 is essentially unknown and our choices of the ranges of some key parameters are mainly based on recent numerical simulations that may be shaped in the future. In view of these facts, the uncertainties of our current estimate of $M_{\rm TOV}$ are likely underestimated. The situation, however, is expected to change considerably in the next decade. With the DNS merger rate of $\sim 10^{3}~{\rm Gpc^{-3}~yr^{-1}}$ (e.g., \citep{Abbott2017a,Jin2018}), the detection rate of the advanced LIGO/Virgo in the full sensitivity run is $\sim 50~{\rm yr^{-1}}$. For such a high detection rate, we can ``optimistically" expect more DNS merger events associated with electromagnetic counterparts thoroughly observed, for example, a gravitational wave event with bright on-axis GRB, which will well determine the mass and angular momentum of torus and ejecta, and improve the estimate of $M_{\rm TOV}$ significantly. Certainly, a robust determination of $M_{\rm TOV}$ still needs the thorough understanding the evolution of merger and the properties of remnant. This task can be achieved by the next-generation missions with the successful detection of postmerger gravitational wave radiation and by the advanced numerical simulations.

\section*{ACKNOWLEDGMENTS}
We thank the anonymous referee for a very helpful suggestion.
This work was supported in part by NSFC under Grants No. 11525313 (the National Natural Fund for Distinguished Young Scholars), No. 11921003, and No. 11433009, by the Chinese Academy of Sciences via the Strategic Priority Research Program (No. XDB09000000), and by the External Cooperation Program of Bureau of International Cooperation (BIC) (No. 114332KYSB20160007). X.S. thanks PMO for the hospitality during his visits.

\appendix
\section{Universal Relations}
Here we follow \citet{Breu2016} to establish some universal relations for both nonrotating and rotating neutron stars.
The EoSs used in our study are not exactly the same as that of \citet{Breu2016}: in addition to the request of $M_{\rm TOV}>2M_\odot$ \citep{Demorest2010, Antoniadis2013, Cromartie2019}, we further apply the condition of $R_{1.4}\leq 14$ km, as inferred from the GW170817 data \citep{Abbott2017a}. Our set of EOSs include AP4, APR4\_EPP \citep{Akmal1998, 2010PhRvC..81a5803T,2009PhRvD..79l4032R, 2016CQGra..33p4001E}; WFF1, WFF2 \citep{1988PhRvC..38.1010W}; RS, SK255, SK272, SKA, SKB, SKI2, SKI3, SKI4, SKI6, SKMP \citep{2015PhRvC..92e5803G, 2009NuPhA.818...36D, 2003PhRvC..68c1304A, 1996PhRvC..53..740N, 1995NuPhA.584..467R, 1989PhRvC..40.2834B, 1986PhRvC..33..335F}; SLY, SLY2, SLY4, SLY9, SLY230A \citep{2001A&A...380..151D, 2015PhRvC..92e5803G, 2009NuPhA.818...36D, 1997NuPhA.627..710C}; HQC18 \citep{2018RPPh...81e6902B}; MPA1 \citep{1987PhLB..199..469M}; ENG \citep{1996ApJ...469..794E}; BSK20, BSK21 \citep{2010PhRvC..82c5804G, 2013A&A...560A..48P}; DD2Y \citep{2017PhRvC..96d5806M, 2010PhRvC..81a5803T}; HS\_DD2, SFHo, SFHx \citep{2004NuPhA.732...24G, 2010NuPhA.837..210H, 2010PhRvC..81a5803T}, LS220 \citep{1991NuPhA.535..331L}. All of the macroscopic properties of a neutron star are calculated by the open-source code RNS \citep{1995ApJ...444..306S, 1992ApJ...398..203C, 1989MNRAS.237..355K} with tables taken from LALSuite built-in EoS data,\footnote{\url{https://git.ligo.org/lscsoft/lalsuite}} EoS catalog\footnote{\url{http://xtreme.as.arizona.edu/NeutronStars/data/eos\_tables.tar}} provided by \citet{Ozel2016}, and online service CompOSE.\footnote{\url{https://compose.obspm.fr/home}} And the quantities at the turning point, e.g., $M_{\rm crit}$, are calculated along a constant angular momentum by increasing central energy density $\varepsilon_{\rm c}$ until
\begin{equation}
\label{eq:turningpoint}
\frac{\partial M(\varepsilon_{\rm c},J)}{\partial \varepsilon_{\rm c}}\Bigg\vert_{J={\rm constant}}\leq 0.
\end{equation}

\begin{figure}
\centering
\subfigure{
\includegraphics[width=0.45\columnwidth]{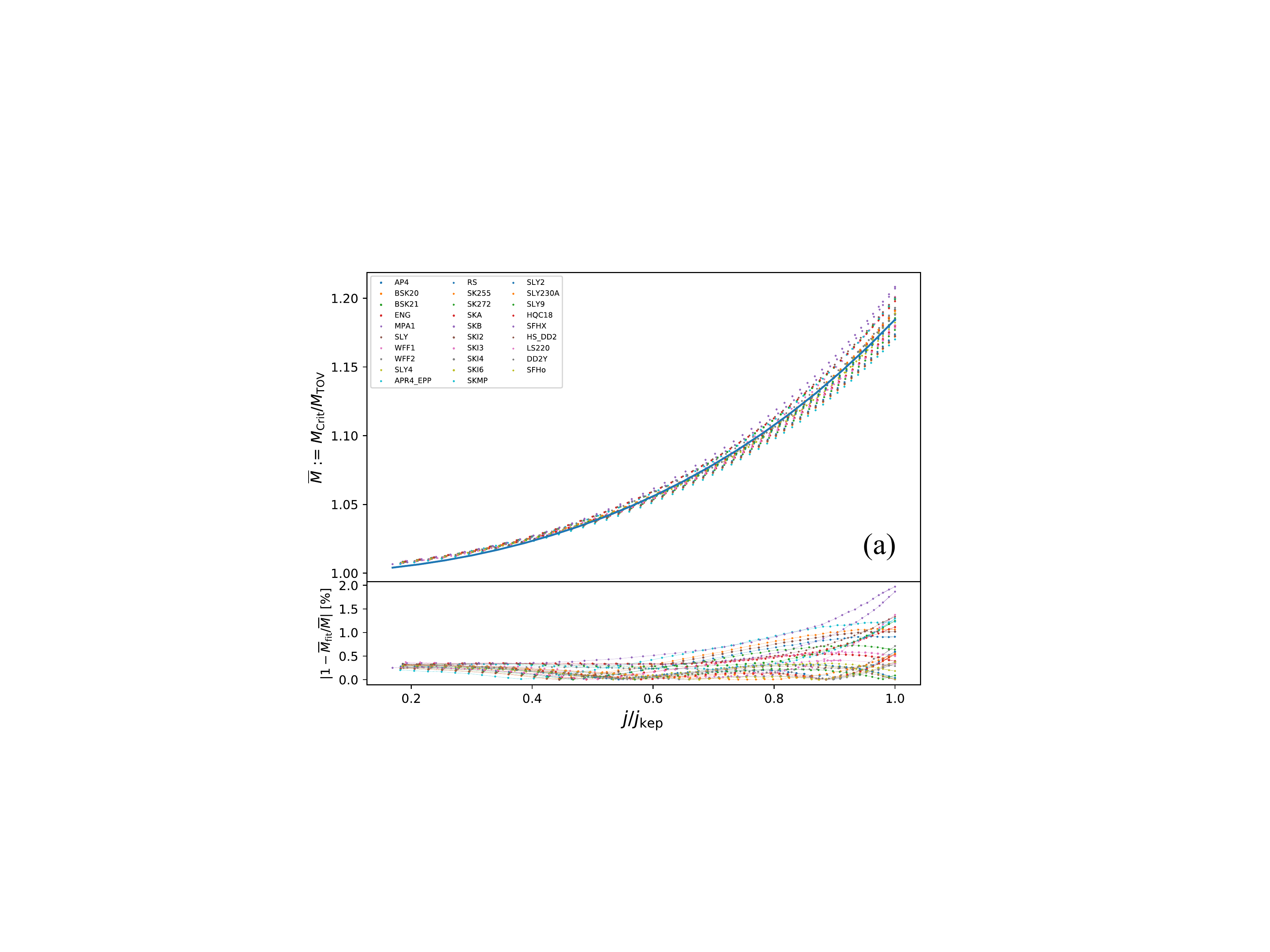}
\label{fig:fit_joverjkep}
}
\hspace{.1in}
\subfigure{
\includegraphics[width=0.45\columnwidth]{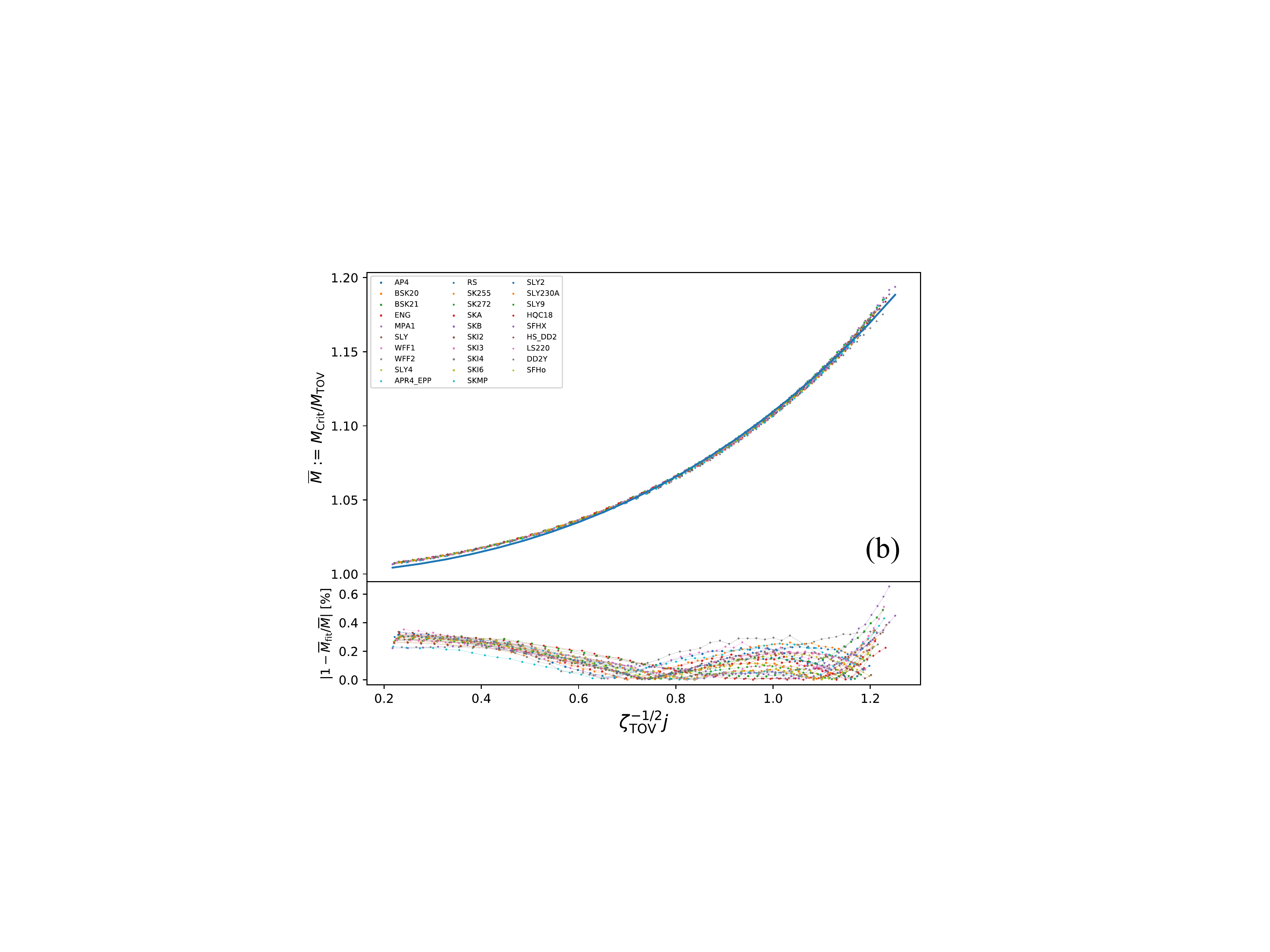}
\label{fig:fit_Mcrit}
}
\caption{Left panel (a): The quantities of $M_{\rm crit}/M_{\rm TOV}$ vs. $j/j_{\rm Kep}$ for a set of EoSs. Shown with a blue solid line is the polynomial fit of Eq.(\ref{eq:McritoverMTOV}). Right panel (b): The quantities of $M_{\rm crit}/M_{\rm TOV}$ vs. $\zeta_{\rm TOV}^{-1/2}j$ calculated with different EoS. Shown with a blue solid line is the polynomial fit of Eq.(\ref{eq:Mcrit}). And the lower panel shows the relative deviation of the numerical data from the fitting function.}
\end{figure}

Similar to Eq.(12) of \citet{Breu2016}, we also find a tight correlation (see Fig.\ref{fig:fit_joverjkep}) of
\begin{equation}
\label{eq:McritoverMTOV}
\frac{M_{\rm crit}}{M_{\rm TOV}}=1+a_2{\left(\frac{j}{j_{\rm Kep}}\right)}^2+a_4{\left(\frac{j}{j_{\rm Kep}}\right)}^4,
\end{equation}
where the best fit coefficients are $a_2=0.1390(9)$, $a_4=0.0455(12)$, with a reduced chi square of $\chi_{\rm red}^2=2.0\times{10}^{-5}$.
\citet{Breu2016} then adopted the approximation of $I_{\rm Kep} \propto M_{\rm Kep}R_{\rm Kep}^2$ to derive their Eq.(17), $j_{\rm Kep}\propto \zeta_{\rm TOV}^{-0.5}$, and Eq.(18), $M_{\rm crit}=(1+c_2\zeta_{\rm TOV}j^2+c_4\zeta_{\rm TOV}^2j^4)M_{\rm TOV}$. However, in reality $I_{\rm Kep}$ is not simply a linear function of $M_{\rm Kep}R_{\rm Kep}^2$ (one can see this, for example, in their Fig.3). Very recently, \citet{2019arXiv190713375K} found a universal relation $I_{\rm Kep}=(-0.006+1.379\zeta_{\rm Kep})M_{\rm Kep}R_{\rm Kep}^2\approx 1.379\zeta_{\rm Kep}M_{\rm Kep}R_{\rm Kep}^2$ (in most cases we have $\zeta_{\rm Kep}\sim 0.27$, and its product with 1.379 is much larger than $-0.006$). Together with the well-established correlations of $M_{\rm Kep}\propto M_{\rm TOV}$, $R_{\rm Kep}\propto R_{\rm TOV}$, and $\Omega_{\rm Kep} \propto \sqrt{M_{\rm Kep}/R_{\rm Kep}^3}$, we have
\begin{equation}
\label{eq:j_Kep}
j_{\rm Kep} = \frac{J_{\rm Kep}}{M_{\rm Kep}^2}
\propto \zeta_{\rm TOV}^{0.5}.
\end{equation}

Therefore we suggest a universal relation of
\begin{equation}
\label{eq:Mcrit}
M_{\rm crit}=(1+c_2\zeta_{\rm TOV}^{-1}j^2+c_4\zeta_{\rm TOV}^{-2}j^4)M_{\rm TOV},
\end{equation}
and the best fit coefficients are $c_2=0.0902(2)$ and $c_4=0.0193(2)$. The reduced chi square is $\chi_{\rm red}^2=3.3\times10^{-6}$. The accuracy of this fit is better than 1\% in the whole range [see Fig.\ref{fig:fit_Mcrit}]. With the best fit values of $c_1$, $c_2$, $c_3$ and $c_4$, it is found that $j_{\rm Kep} \approx 1.24~\zeta_{\rm TOV}^{0.5}$.

\begin{figure}
\centering
\includegraphics[width=0.8\textwidth]{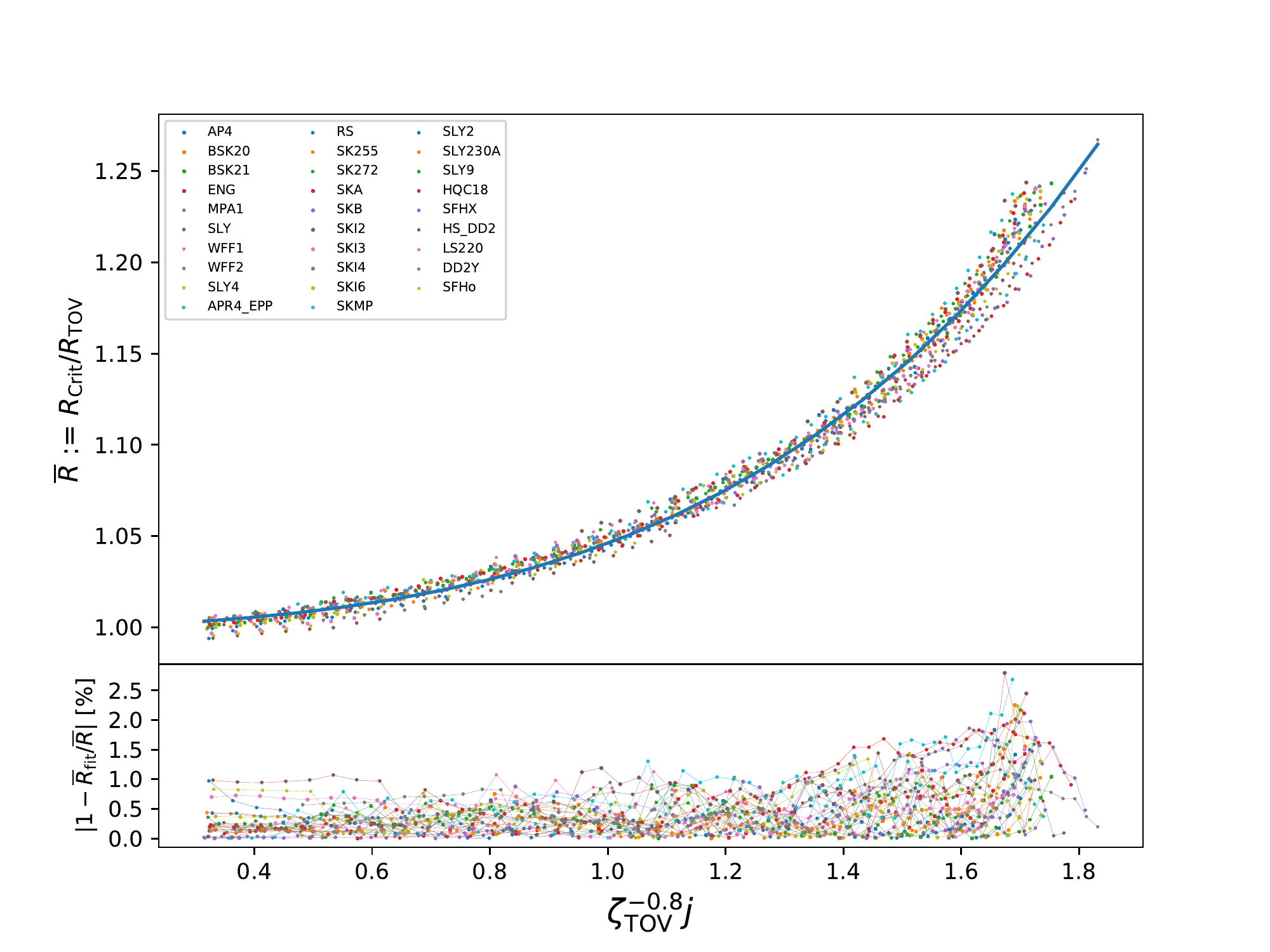}
\caption{The quantities of $R_{\rm crit}/R_{\rm TOV}$ vs. $\zeta_{\rm TOV}^{-0.8}j$ calculated with different EoS. Shown with a blue solid line is the polynomial fit of Eq.(\ref{eq:Rcrit}), while the lower panel shows the relative deviation of the numerical data from the fitting function.}
\label{fig:fit_Rcrit}
\end{figure}
Another relation necessary for this work is the equatorial radius of neutron star ($R_{\rm crit}$) corresponding to the critical mass $M_{\rm crit}$ as a function of $\zeta_{\rm TOV}$, $j$ and $R_{\rm TOV}$. The empirical function reads
\begin{equation}
\label{eq:Rcrit}
R_{\rm crit}=(1+d_2\zeta_{\rm TOV}^{-1.6}j^2+d_4\zeta_{\rm TOV}^{-3.2}j^4)R_{\rm TOV},
\end{equation}
where the best fit coefficients are $d_2=0.0321(5)$, $d_4=0.0139(2)$, with a reduced chi square of $\chi_{\rm red}^2=4.9\times{10}^{-5}$. The accuracy of this fit is better than a few percent, as shown in Fig.\ref{fig:fit_Rcrit}.

\begin{figure}
\centering
\includegraphics[width=0.8\columnwidth]{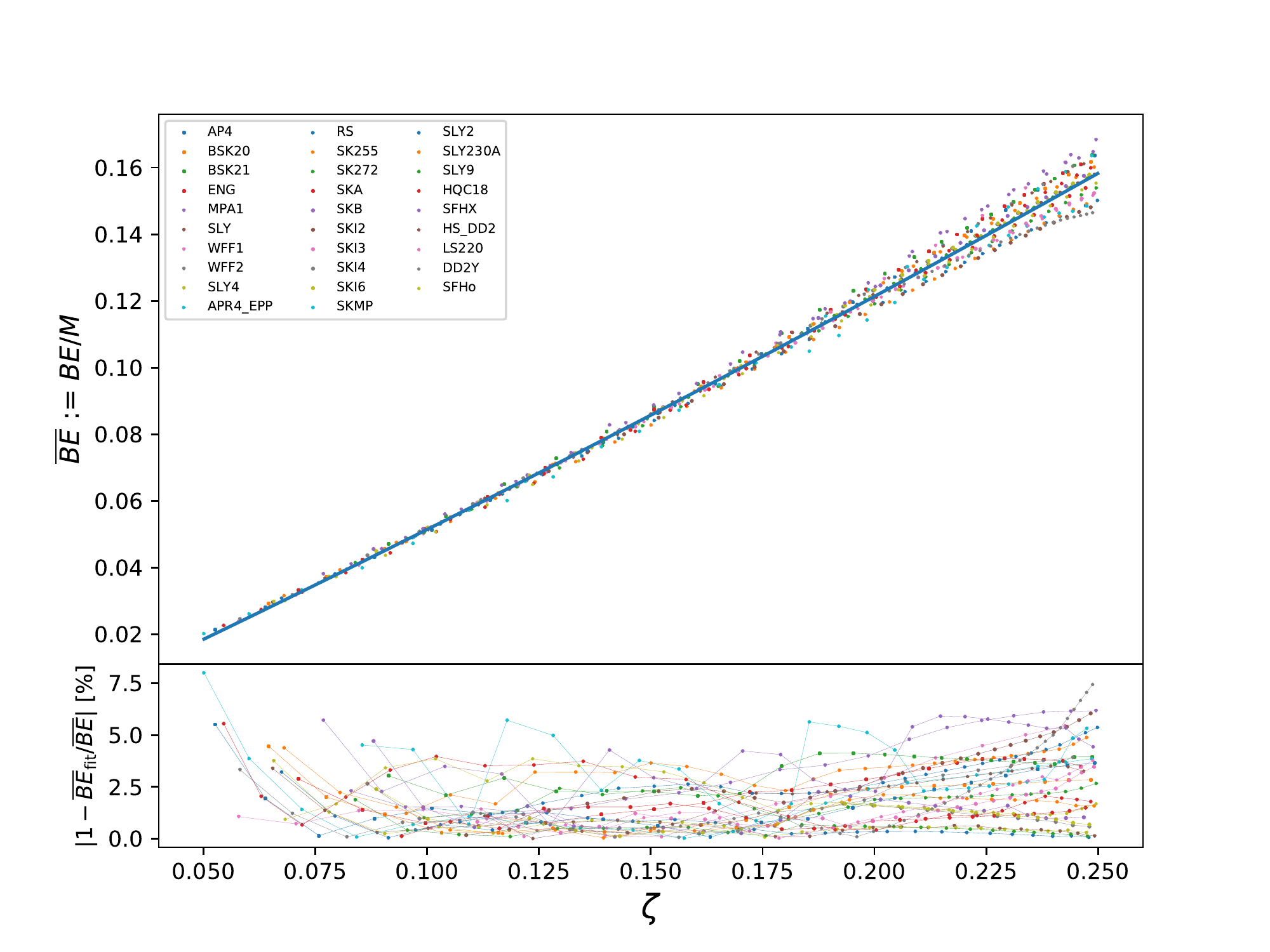}
\caption{The quantities of $BE/M$ vs $\zeta$ calculated with different EoSs. Shown with a blue solid line is the polynomial fit $BE/M=b_0+b_1\zeta+b_2\zeta^2$ with the coefficients of $b_0=-0.0130(12), b_1=0.618(16)$, $b_2=0.267(48)$ and $\chi_{\rm red}^2=7.8\times{10}^{-5}$, while the lower panel shows the relative deviation of the numerical data from the fitting function.}
\label{fig:fit_BE}
\end{figure}

\begin{figure}
\centering
\subfigure{
\includegraphics[width=0.45\columnwidth]{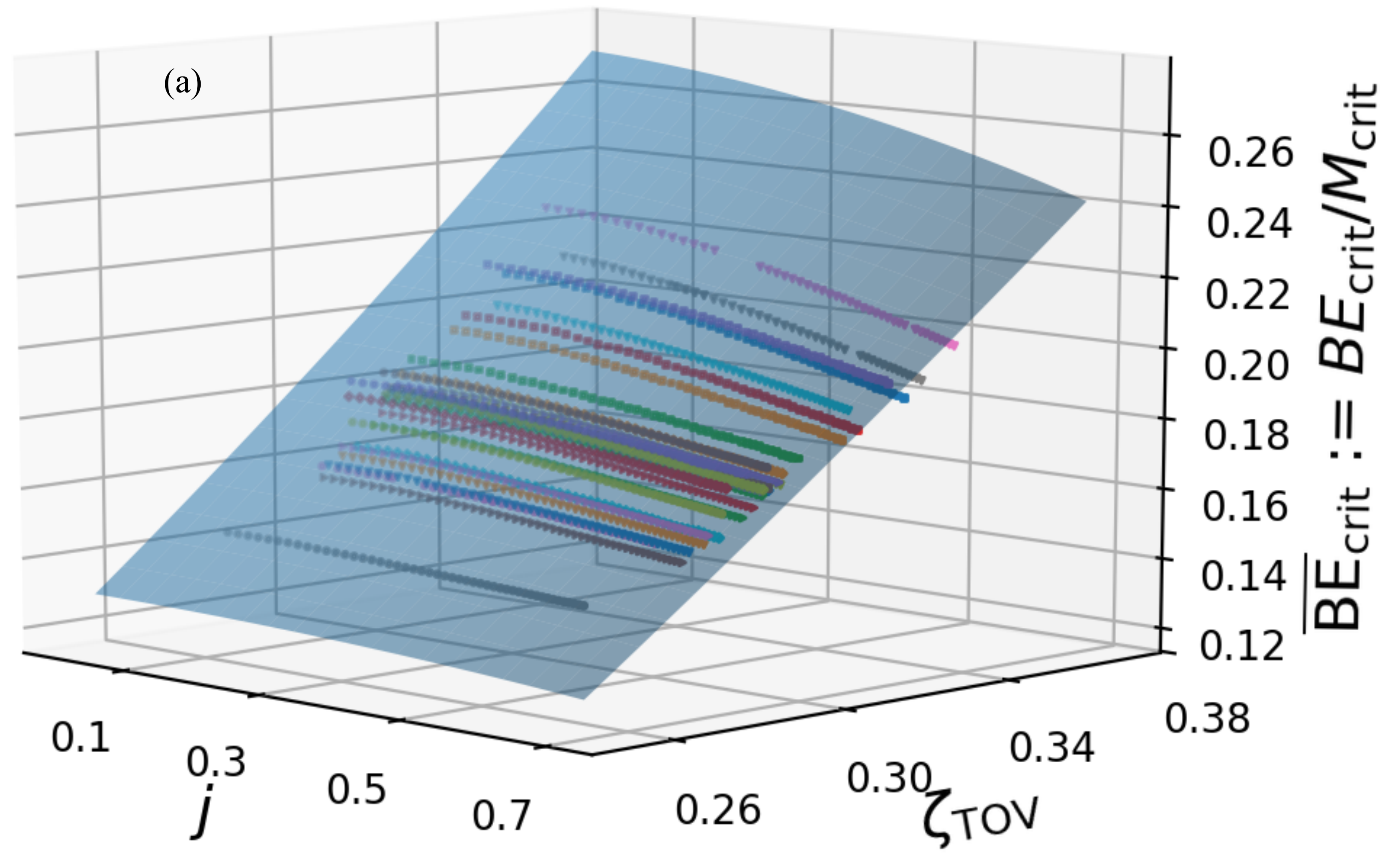}
\label{fig:fit_BEcrit}
}
\hspace{.1in}
\subfigure{
\includegraphics[width=0.45\columnwidth]{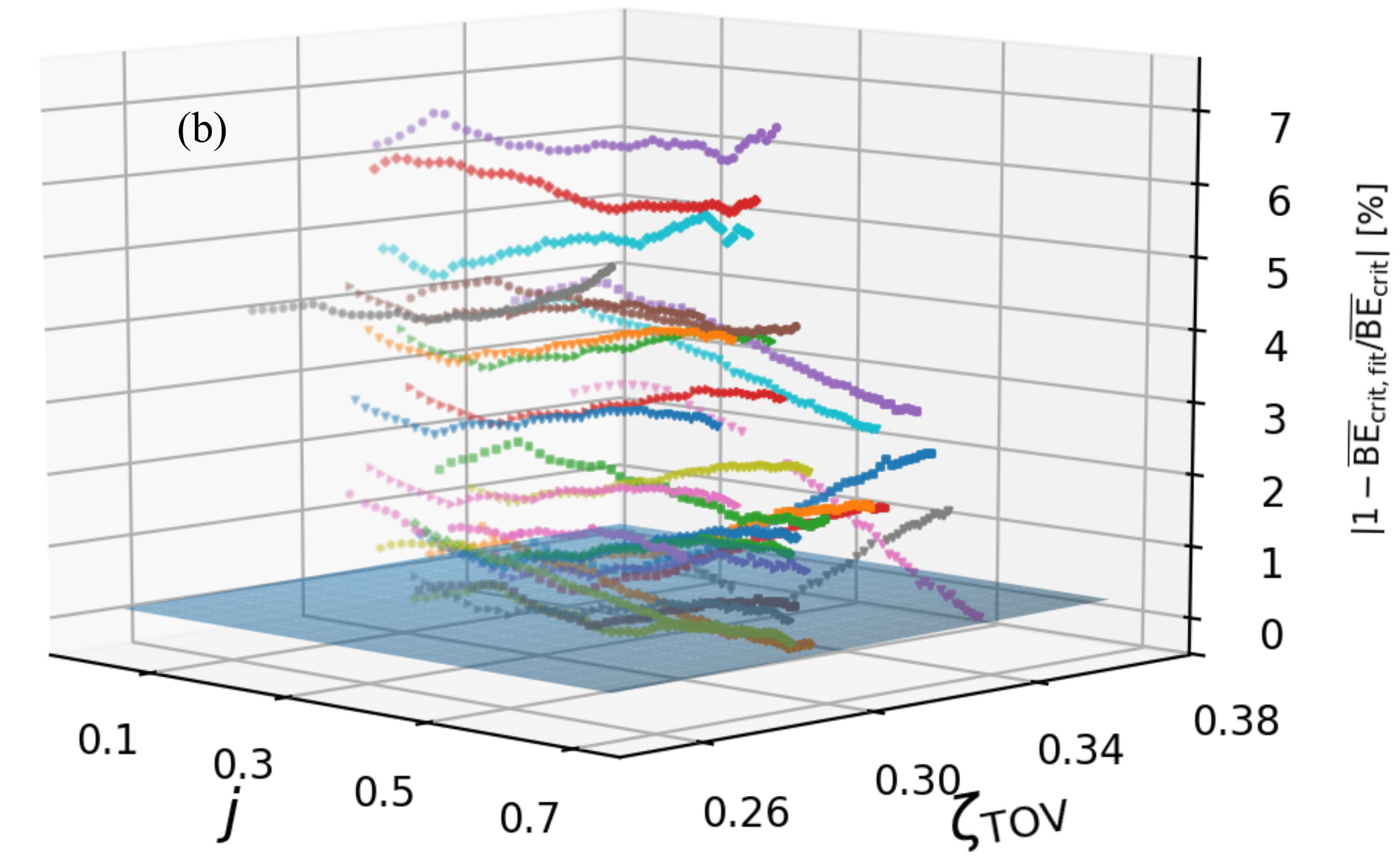}
\label{fig:deviation_fit_Mcrit}
}
\caption{Left panel (a): The quantities of $BE_{\rm crit}/M_{\rm crit}$ vs. $\zeta_{\rm TOV}$ and $j$ calculated with different EoSs. Shown with a blue surface is the fit of Eq.(\ref{eq:BEcrit}). Right panel (b): The relative deviation of each EoS from the fitted value.}
\end{figure}

Meanwhile, we also get a universal relation of a nonrotating neutron star between dimensionless binding energy $BE/M$ and compactness $\zeta$, which are similar to that reported in the literature \citep[e.g.,][]{Breu2016, 2016EPJA...52...18S}. As shown in Fig.\ref{fig:fit_BE}, for $0.05<\zeta<0.25$, the correlations among these factors are tight.
We also fit $BE_{\rm crit}/M_{\rm crit}$ as a function of $j$ and $\zeta_{\rm TOV}$, which takes the form of
\begin{equation}
\label{eq:BEcrit}
\frac{BE_{\rm crit}}{M_{\rm crit}}=e_0+e_1(1+\alpha j+\beta j^2)\zeta_{\rm TOV}+e_2(1+\gamma j+\delta j^2)\zeta_{\rm TOV}^2,
\end{equation}
where the best fit coefficients are $e_0=-0.10$, $e_1=0.78$, $\alpha=-0.050$, $\beta=-0.034$, $e_2=0.61$, $\gamma=0.23$ and $\delta=-0.58$, and the reduced chi square is $\chi_{\rm red}^2=1.9\times10^{-4}$ [see Fig.\ref{fig:fit_BEcrit} for the fit results].

\clearpage

\end{document}